\documentclass[aps,prl,twocolumn,showpacs,superscriptaddress]{revtex4}
\usepackage{graphicx}

\begin{document}
\title{Exchange Interaction between Single Magnetic Adatoms}
\author{P. Wahl}
\affiliation{Max-Planck-Institut f\"ur Festk\"orperforschung,
Heisenbergstr. 1, 70569 Stuttgart, Germany}%
\author{P. Simon}
\affiliation{Laboratoire de Physique et Mod\'{e}lisation des Milieux
Condens\'{e}s, Universit\'{e} Joseph Fourier and CNRS, 25 av. des
Martyrs, F-38042 Grenoble, France}
\author{L. Diekh\"oner}
\affiliation{Institut for Fysik og Nanoteknologi, Aalborg
Universitet, Skjernvej 4, DK-9220 Aalborg, Denmark}
\author{V.S. Stepanyuk}
\author{P. Bruno}
\affiliation{Max-Planck-Institut f\"ur Mikrostrukturphysik, Weinberg
2, D-06120 Halle, Germany}
\author{M. A. Schneider}
\affiliation{Max-Planck-Institut f\"ur Festk\"orperforschung,
Heisenbergstr. 1, 70569 Stuttgart, Germany}
\author{K. Kern}
\affiliation{Max-Planck-Institut f\"ur Festk\"orperforschung,
Heisenbergstr. 1, 70569 Stuttgart, Germany}
\affiliation{Institute de  Physiques des Nanostructures, Ecole
Polytechnique F\'{e}d\'{e}rale de Lausanne, CH-1015 Lausanne,
Switzerland}
\date{\today}

\begin{abstract}
The magnetic coupling between single Co atoms adsorbed on a copper
surface is determined by probing the Kondo resonance using
low-temperature scanning tunneling spectroscopy. The Kondo
resonance, which is due to magnetic correlation effects between the
spin of a magnetic adatom and the conduction electrons of the
substrate, is modified in a characteristic way by the coupling of
the neighboring adatom spins. Increasing the interatomic distance
of a Cobalt dimer from 2.56 \AA{} to 8.1 \AA{} we follow the
oscillatory transition from ferromagnetic to antiferromagnetic coupling.
Adding a third atom to the antiferromagnetically coupled dimer results in the formation of a collective correlated state.
\end{abstract}

\pacs{72.10.Fk, 72.15.Qm, 68.37.Ef}

\maketitle

The magnetic properties of nanostructures play a pivotal role in the
design of miniaturized spin-based devices. One of the key parameters
is the magnetic interaction between the constituent atoms of a
nanostructure. This interaction can be due to direct or indirect
coupling as well as mediated via a supporting substrate or host.
Depending on the strength and sign of the exchange interaction, the
nanostructure can be driven into ferromagnetic or antiferromagnetic
behaviour, a correlated state or complex spin structures
\cite{Asada06}. However, up to now it was impossible to measure
experimentally the magnetic interaction between individual atoms. Only recently spin-flip experiments by Hirjibehedin and coworkers \cite{Hirjibehedin06} have enabled a direct probing of the magnetic interaction in linear manganese chains decoupled from the substrate by a spacer layer. In the present letter
we exploit the Kondo effect as a local probe to determine the
exchange interaction between individual cobalt adatoms on a metallic
substrate as a function of their distance. This approach was originally proposed by Chen {\it et al.} \cite{Chen99}. Their experiment, however, did not allow for an assessment of the exchange coupling, because for the studied nearest neighbor dimers on Au(111) only the disappearance of the Kondo resonance was observed.

The Kondo effect originates from the screening of the spin of a
magnetic impurity by the surrounding conduction band electrons
\cite{Hewson93} and is characterized by a strong peak in the
impurity's density of states near the Fermi level. This many body
resonance has been observed for single magnetic adatoms
\cite{Madhavan98,Li98,Knorr02}, in artificial nanostructures such as
quantum corrals \cite{Manoharan00} and for molecules
\cite{Zhao05,Wahl05}. In scanning tunneling spectroscopy (STS) spectra, it shows up as a feature which
can be described by a Fano line shape
\cite{Fano61,Plihal01,Ujsaghy00}. From a fit, the peak width
$\Gamma$ is obtained which is the characteristic energy scale -- the
Kondo temperature $T_{\mathrm K}$ -- of the impurity system. For the
Kondo scenario of a single magnetic impurity on a nonmagnetic metal
surface a quantitative description has been proposed \cite{Wahl04}.

\begin{figure*}[hbt]
\includegraphics[width=16 cm]{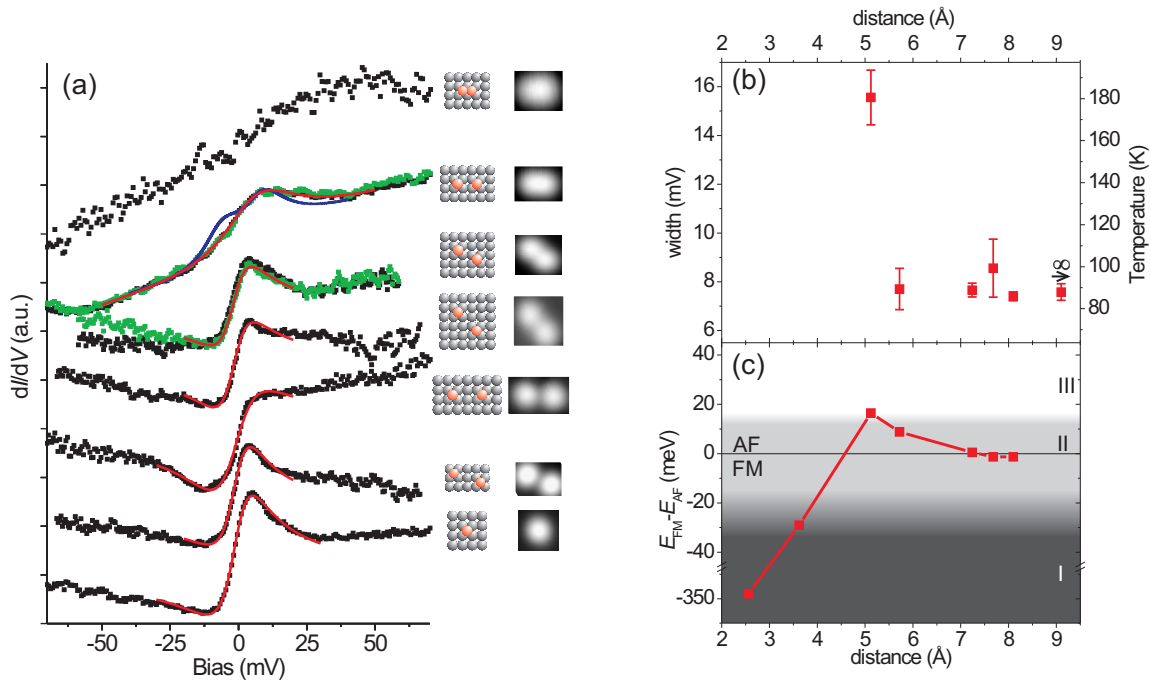}
\caption[]{Kondo resonance of cobalt dimers on Cu(100) measured by
STS at 6~K. As a consistency check,
spectra taken on both ends of the dimers are shown (green and black
dots) to be equivalent. (a) Model, topography and spectra for (from top to bottom) a compact dimer (2.56~\AA), a dimer at 5.12~\AA, at 5.72~\AA, at 7.24~\AA, at 7.68~\AA, at 8.10~\AA~and for a single adatom at infinite distance ($>$ 20~\AA) are depicted. The spectra are shown together with fits of a Fano function (red solid line), for the dimer at 5.12~\AA also a simulated curve according to Eq. \ref{eq1} with $J = 15$~meV and $\Gamma = 1.2 T_{\mathrm K}^0$ is plotted (blue solid line). For the dimer at 5.12~\AA, a linear background had to be taken into account to obtain a reasonable fit for a Fano function. (spectra shifted vertically for clarity) (b) shows the width of the resonance as a function of distance and (c) KKR calculations for the exchange interaction between cobalt adatoms on Cu(100) \cite{Stepanyuk01}. The three distinct regimes discussed in the text are shaded in different greys.}
\label{fig-wahl1}
\end{figure*}
As a second impurity is brought into proximity, magnetic
interactions between the impurities become important and may
modify the Kondo resonance considerably. These magnetic interactions
can be of different origin: magnetic dipolar coupling, direct
exchange between the impurities due to an overlap of the impurity
$d$ orbitals, or finally the Ruderman-Kittel-Kasuya-Yosida (RKKY)
interaction. The latter is an indirect spin-spin interaction
mediated by the conduction electrons of the host. Here, we demonstrate that it is possible to determine the magnetic
interaction between single magnetic atoms adsorbed on a noble metal
surface by measuring the modified Kondo spectrum. The results are compared to theoretical predictions of the magnetic interactions between single atoms \cite{Stepanyuk01}. The evolution of the
Kondo line shape obtained by STS upon varying the interatomic
distance between Co adatoms in dimer and trimer configurations on a
Cu(100) single crystal surface is compared to many body theory,
which allows us to determine the magnetic coupling as function of
the interatomic spacing.
Since manipulation of atoms on (100) metal surfaces is difficult due
to a high energy barrier between neighboring sites, dimers with
well-defined interatomic distance have been fabricated with an
alternative approach. First, cobalt carbonyl complexes are formed
\cite{Wahl05}, in which the CO ligands inhibit nucleation and island
formation and facilitate the growth of one-dimensional structures.
Once these structures are grown, the CO ligands can be removed by
tip-induced dissociation of the molecules leaving behind on the
surface only the cobalt atoms. The panels in
Fig.~\ref{fig-wahl1}(a) show different cobalt dimer configurations
prepared as described above with interatomic distances between the
neighboring cobalt atoms ranging from 2.56~\AA{} to 8.1~\AA{}
together with the corresponding STS spectra.
For the compact dimer the interaction
between the spins is much stronger than the coupling to the
substrate and the Kondo effect (at 6~K) is suppressed. For the
next-nearest neighbor distance, however, a resonance is found at
the Fermi energy. The resonance is considerably broader than that of a single cobalt adatom. By fitting
the STS signal with a single Fano line shape, we extract that the
energy width of the feature would correspond to a Kondo temperature
$T_{\mathrm K} = 181 \pm 13$~K. For distances of 5.72~\AA{} and
7.24~\AA{}, the Kondo resonance has
already recovered almost the same width and line shape as that of a
single cobalt adatom.
For even larger distances, the same width as on an adatom is restored. The widths of the resonances are tabulated in Table~\ref{tab-wahl1} and summarized in Fig.~\ref{fig-wahl1}(b).
\begin{table}[hbt]
\begin{tabular}{lllll}
\hline\hline  & $d$(\AA)& Width (meV) & $\varepsilon_{\mathrm K}$
(meV) &
$q$ \\ \hline %
Monomer & $-$ & $7.58 \pm 0.34$ & $-1.3 \pm 0.4$ & $1.13 \pm 0.06$ \\
Dimer & 2.56 & $-$ & $-$ & $-$ \\
Dimer & 5.12 & $15.6 \pm 1.1$ & $5.3 \pm 1.6$ & $3.2 \pm 1.1$ \\
Dimer & 5.72 & $7.71 \pm 0.85$ & $-0.95 \pm 0.73$ & $1.56 \pm 1.04$ \\
Dimer & 7.24 & $7.66 \pm 0.28$ & $-0.93 \pm 0.70$ & $1.17 \pm 0.14$ \\
Dimer & 7.68 & $8.89 \pm 1.57$ & $-0.94 \pm 1.20$ & $1.10 \pm 0.48$ \\
Dimer & 8.1  & $7.40 \pm 0.18$ & $-0.72 \pm 0.14$ & $1.44 \pm 0.04$ \\
Trimer& 5.12 & $13.23 \pm 0.91$ & $-1.43 \pm 0.76$ & $27.2 \pm 0.3$ \\
      &  & $11.93 \pm 1.15$ & $-0.69 \pm 0.72$ & $-0.03 \pm 0.01 $ \\
\hline\hline
\end{tabular}
\caption[]{Width, position $\varepsilon_{\mathrm K}$ and line shape
parameter $q$ of a single Fano resonance fit to the feature of Cobalt adatoms, dimers and
trimers on Cu(100) measured by STS at 6~K. For dimers and trimers,
the interatomic distances d and for the trimer the data for both
features at the Fermi energy are given (Monomer value taken from
Ref.~\cite{Knorr02}). The errors given are the standard deviation of averages over measurements taken on different adatoms pairs and with different tips. } \label{tab-wahl1}
\end{table}

Our data can be theoretically interpreted as a realization of a
two-impurity Kondo problem \cite{Jayaprakash81}. Depending on the
relative strength of the exchange interaction compared to the single
impurity Kondo temperature $T_{\mathrm K}^0$, the dimers enter
different regimes. For a strong ferromagnetic exchange interaction
$|J| \gg T_{\mathrm K}^0$ [marked as regime I in
Fig.~\ref{fig-wahl1}(c)] a correlated state with a new Kondo
temperature $T_{\mathrm K}^\mathrm{dimer} \approx \left(T_{\mathrm
K}^0\right)^2/|J|$ will occur \cite{Jayaprakash81}. This new Kondo
scale is much lower than the temperature of the experiment and can
therefore not be detected in our measurements. For intermediate
exchange interaction $J$ [regime II in Fig.~\ref{fig-wahl1}(c)], the
single impurity Kondo resonance is recovered. Finally, for a
sufficiently strong antiferromagnetic exchange interaction $J > J^*
\sim 2 T_{\mathrm K}^0$ [marked as regime III in
Fig.~\ref{fig-wahl1}(c)] between neighboring magnetic atoms, the
Kondo resonance is split and a singlet state is formed between the
impurities \cite{Jones87}. This singlet state is characterized in
the impurity density of states by peaks located at energies $\pm
J/2$ \cite{Lopez02,Simon05,Vavilov05}. The splitting of the Kondo
resonance for strong antiferromagnetic coupling has been observed
previously in quantum dots coupled either by a direct tunnel
junction \cite{Jeong01} or by an indirect RKKY interaction mediated
via another large quantum dot \cite{Craig04}. For adatoms on a
substrate, we can show using a slave boson mean field
theory approach as in Ref.~\onlinecite{Simon05} that this splitting generates the following
density of states:
\begin{equation}
\rho(\varepsilon) \propto a_1 f\left( \frac{\varepsilon +
J/2}{\Gamma},q\right) + a_2 f \left(\frac{\varepsilon -
J/2}{\Gamma},q\right), \label{eq1} \end{equation} %
where $f(x,q) = \frac{\left(x+q\right)^2}{x^2 +1}$
and $a_1 \sim a_2$. This equation describes two Fano resonances at
$\pm J/2$  with a width $\Gamma$ which is of the same order as the
single impurity one. The resonances are resolved in the tunneling
spectrum as only one broadened feature as observed in the experiment
due to the width of the resonances, which is of the same order as
the splitting. Thus the width of the resonance in this case provides
a measure for the magnetic interaction between the adatoms.

In Table~\ref{tab-wahl1}, the experimental results for the atomic
arrangements which have been investigated are summarized. For the
compact dimer (2.56~\AA), we find that the Kondo resonance disappears.
This is consistent both with previous experiments on Co dimers on Au(111)
by Chen {\it et al.} \cite{Chen99} and the strong ferromagnetic coupling predicted by
{\it ab initio} calculations \cite{Stepanyuk01}, which introduces a
Kondo scale $T_{\mathrm K}^\mathrm{dimer} \sim 2$~K for the nearest neighbour dimer, which is
smaller than the temperature of the experiment. The spectrum on the
next-nearest neighbor dimer (5.12~\AA) shows a distinct Kondo
resonance at the Fermi level, which is broadened compared to the
spectrum of the isolated Co adatom. By using Eq.~(\ref{eq1}) to fit
the experimental data, we extract an antiferromagnetic coupling $J$
of about 16~meV. The broadened spectrum can be rationalized by an
antiferromagnetic coupling between the two Co adatoms. The relevant
Kondo energy scale is $k_{\mathrm B}T_{\mathrm K} =7.58$~meV for a
Co atom on Cu(100) \cite{Stepanyuk01}. The magnetic interactions are
thus large enough to induce a splitting of the Kondo effect but
sufficiently small to prevent complete quenching as observed for the
compact dimer. At larger interatomic distances the spectrum and
$T_{\mathrm K}$ transforms back to the single adatom value, with the
exception of an interatomic distance of 7.68~\AA, where the
resonance width has a local maximum as a function of distance.
According to {\it ab initio} calculations \cite{Stepanyuk01}, the
interaction between two cobalt adatoms on Cu(100) is mainly due to
RKKY interactions. When the adatoms are on next-nearest neighbor
sites, Stepanuyk {\it et al.} predict an antiferromagnetic
interaction of about 17~meV. This is in excellent agreement with the
estimation we obtain assuming a split Kondo resonance. When the
adatoms are further apart (5.72~\AA), the calculations predict that
the RKKY interaction is reduced to 8~meV, which is not large enough
to split the Kondo resonance and therefore explains why the usual
single-impurity Kondo resonance is almost recovered.

\begin{figure}[hbt]
\includegraphics[width=8 cm]{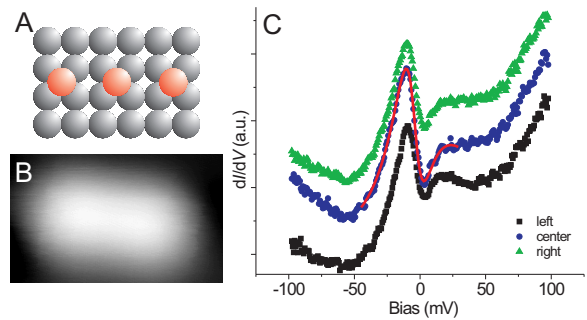}
\caption[]{Kondo resonance of the Cobalt trimer on Cu(100) measured
by STS at 6~K. (a) Model of the trimer
investigated, (b) STM topography (same scale as model). (c) Spectra
taken on the left, right and center atom, the spectra are shifted
vertically. The solid line is a fit to a double Fano resonance.}
\label{fig-wahl2}
\end{figure}

Besides cobalt dimers, we studied a linear trimer. Trimers of magnetic adatoms have been studied previously by Jamneala {\it et al.} \cite{Jamneala01}. Their study focussed on compact Cr trimers, no statement was made on the low energy excitations of a linear trimer. Our trimer has the
adatoms on next-nearest neighbor sites as shown in Fig.~\ref{fig-wahl2}(a) and (b).
The trimer has been prepared in a similar way as the dimers. The
tunneling spectra change qualitatively on the trimer. As can be seen
from Fig.~\ref{fig-wahl2}(c), the spectra show a superposition of
features -- resulting in two maxima and a dip in between them. %

In order to understand the STS spectra for the linear trimer, we
study theoretically within a slave boson mean field theory (SBMFT)
\cite{Asada06} and perturbative renormalisation group (PRG) theory
\cite{Lazarovits05} a cluster of three inline magnetic impurities
coupled to a conduction band. As for the dimer, two scenarios are
possible: Either the RKKY interaction is not strong enough to split
the Kondo resonances ($J \sim 2 T_{\mathrm K}^0$) and the middle
impurity and side impurities have their own Kondo resonances of
different widths or the RKKY interaction is sufficiently strong to
induce a collective magnetic behaviour ($J \stackrel{>}{\sim} 2
T_{\mathrm K}^0$). In the latter case, the SBMFT predicts three Fano
features centered at $e V = 0, \pm J/2$ . Additionally, for strong
antiferromagnetic interaction the ground state has a net spin which
is screened at low temperatures giving therefore rise to an extra
Kondo resonance around zero bias. This extra Kondo resonance is not
captured by SBMFT. An analysis similar to the one developed by
Lazarovits {\it et al.} \cite{Lazarovits05} based on PRG indicates
that the width of this resonance can be of order of $T_{\mathrm
K}^0$ or even larger.

In the first case, the STS signal is a sum of two single-impurity
Fano resonances whose amplitudes depend on the position of the STM
tip along the cluster. In the second case, the STS signal can be
understood as the sum of two Fano resonances centered around zero
bias plus a dimer (split) Fano resonance at $e V = \pm J/2$. Since in
both cases at least two zero-bias features are predicted, we fit our
data with two Fano resonances. The values we obtain are also given
in Table~\ref{tab-wahl1}. In order to discriminate between both
scenarios, we have analyzed the spatial dependence of the relative
amplitude of both zero-bias features. We find a negligible
dependence of the line shape on the tip position especially when
moving the tip away from the chain. This clearly favors the second
scenario where both zero-biased resonances are collective features.
This conclusion is consistent with the calculated strength of the
magnetic interaction in the trimer. We have performed calculations employing the Korringa-Kohn-Rostocker Green's-function method (KKR) (as described in \cite{KKR}) which reveal an energy difference between the ferromagnetic and the
antiferromagnetic configuration of the trimer of 35~meV.
This yields a nearest-neighbor (NN) spin interaction in the trimer of
$J_{\mathrm NN} \sim 17$~meV (comparable to the dimer value), which
is larger than $T_{\mathrm K}^0$ and puts the trimer in the
correlated regime. While for chains consisting of Mn atoms decoupled
from the metallic substrate as studied by Hirjibehedin {\it et al.}
\cite{Hirjibehedin06} collective magnetic excitations have been
observed, the correlated electronic state found here is induced by
the coupling of the magnetic atoms via the substrate.

The situation in the Co chain also differs from the
compact Cr trimer studied by Jamneala {\it et al.} \cite{Jamneala01}
where the magnetic exchange interactions between the adatoms have
been shown to be several orders of magnitude larger than the Kondo
temperature of a single Chromium adatom leading to different
behaviours depending very strongly on the shape of the trimer
\cite{Lazarovits05}.

In conclusion, we have shown how the magnetic
interaction between single magnetic atoms coupled to a substrate can
be determined via the Kondo effect. Understanding and being able to
measure the magnetic coupling on the single atom level is expected
to play a key role in the design of magnetoelectronic devices.
Magnetic nanostructures with specific properties can be tailored by
controlling and manipulating the exchange interaction. Furthermore
the transition from a single-impurity Kondo effect to a collective
state of a small chain or cluster offers a unique opportunity to
compare theories for the 1D Kondo chain to experimental model
systems.

PS has been partially supported by the contract PNANO 'QuSpins' of the french ANR.

\end{document}